\documentclass[10pt,aip,apl,twocolumn,amssymb,floatfix,superscriptaddress]{revtex4-1}
\usepackage{dcolumn,amsmath,bm,epsfig}
\usepackage[colorlinks,linkcolor=blue,citecolor=blue]{hyperref} 
\usepackage{graphicx}

\newcommand{\htc}[1]{\textcolor{black}{#1}}

\begin{document}
\title{Spin Filtering in Germanium/Silicon Core/Shell Nanowires with Pseudo-Helical Gap}
\author{Jian Sun}
 \email[Corresponding author: ]{jian.sun@csu.edu.cn}
 \affiliation{School of Physics and Electronics, Central South University, 932 South Lushan Road, Changsha 410083, China}
 \affiliation{Advanced Device Laboratory, RIKEN, 2-1 Hirosawa, Wako, Saitama 351-0198, Japan}
 
\author{Russell S. Deacon}
  \affiliation{Advanced Device Laboratory, RIKEN, 2-1 Hirosawa, Wako, Saitama 351-0198, Japan}
  \affiliation{Center for Emergent Matter Science, RIKEN, 2-1 Hirosawa, Wako, Saitama 351-0198, Japan}
  
\author{Xiaochi Liu}
 \affiliation{School of Physics and Electronics, Central South University, 932 South Lushan Road, Changsha 410083, China}
  
\author{Jun Yao}
  \affiliation{Department of Electrical and Computer Engineering, Institute for Applied Life Sciences, University of Massachusetts, Amherst, MA 01003, USA}
 
 \author{Koji Ishibashi}
  \affiliation{Advanced Device Laboratory, RIKEN, 2-1 Hirosawa, Wako, Saitama 351-0198, Japan}
  \affiliation{Center for Emergent Matter Science, RIKEN, 2-1 Hirosawa, Wako, Saitama 351-0198, Japan}

\date{\today}

\begin{abstract}
Semiconductors with strong spin-orbit interactions can exhibit a helical gap with spin-momentum locking opened by a magnetic field. Such a gap is highly spin selective as a result of a topologically protected spin-momentum locking, which can be used for spin filtering.
We experimentally demonstrate such a spin filtering effect in a quasi-ballistic $p$-type germanium/silicon core/shell nanowire (NW), which possesses a pseudo-helical gap without the application of magnetic field.
Polarized hole spin injection to the NW is achieved using cobalt ferromagnetic contacts with controlled natural surface oxide on the NW as tunnel barrier. Local and nonlocal spin valve effects are measured as the verification of polarized spin transport in the NW outside the helical gap.
We electrically tune the NW into the helical gap by scanning its chemical potential with a gate. A hysteresis loop with three resistance states is observed in the local spin valve geometry, as an evidence of spin filtering in the helical gap.

\end{abstract}
\maketitle

Germanium/Silicon core/shell nanowires are promising material platform for spintronics. Owing to a large valance band offset of $\sim 0.5\,$eV between Ge and Si, \htc{holes of the concentration of $\sim 10^{17}\,$cm$^{-3}$ are naturally accumulated in the Ge core and strongly confined by the interface with the Si shell.\cite{Xiang-Nature-2006,Lu-PNAS-2005, Li-APL}} The dopant-free growth leads to high mobility with mean-free-path up to $\sim 500\,$nm.\cite{Lu-PNAS-2005} As group IV semiconductors, Ge and Si have a low density of nuclear spins or can be grown with zero net nuclear spin, which may highly suppress the hyperfine coupling induced
spin relaxation. A long spin relaxation in the order of milliseconds has been reported.\cite{Hu-NN-2012}
In addition, the hole system offers several potential advantages for spintronics. Having an effective spin number of $3/2$, hole spin and momentum are strongly coupled to enable electric field mediated spin manipulation. Moreover, hole spin lifetimes can be further prolonged in the presence of confinement \cite{Fischer-PRL,Brunner-science}. 
More importantly, Ge/Si NWs possess a strong dipole-coupled Rashba type spin-orbit interaction as a result of the quasi-degeneracy in its low energy valence bands.\cite{Loss-SOI} 
\htc{In one-dimensional channels possessing strong Rashba spin-orbit interaction, two spin-degenerate subbands are shifted laterally in momentum space, therefore lifting the spin degeneracy. By applying a magnetic field perpendicular to the spin-orbit field, a helical gap is opened at the band touching point, inside of which spin-momentum locking is topologically protected. \cite{Quay-NP-2010} 
By tuning the chemical potential electrostatically using a gate, the transport in the NW can be effectively set inside and out of the helical gap. In one-dimensional ballistic NWs, such a helical gap is detected as a re-entrant conductance feature on quantized conductance plateaux of integer multiples of $2e^2/h$ in transport measurements.\cite{Delft-helical, Julich-helical, SUN-NL} }

The spin-momentum locking allows the helical gap to be employed for spin filtering.
This has been predicted in a different system, \textit{i.e.} topological insulators with edge helical states.\cite{TI-SF-1, TI-SF-2}
So far, we are not aware of any reports on polarized spin transport in the Ge/Si NW.
Such measurements are challenging as fields required to open a helical gap can be of order of a few tesla. 
\htc{Fortunately, in Ge/Si NWs, the strongly correlated two-particle backscattering induces a pseudo-helical gap at zero magnetic field, enabling its potential applications in spintronics at low fields.\cite{Julich-helical, SUN-NL}}

In this study we realize polarized hole spin injection into a $300\,$nm-long quasi-ballistic Ge/Si core/shell NW by using cobalt ferromagnetic (FM) contacts with controlled natural oxide on the NW as a tunnel barrier.
A spin valve effect is measured in both nonlocal and local configurations, revealing the polarized spin transport, when the NW is gated outside the helical gap.
By electrically tuning the transport in the NW inside the pseudo-helical gap with a gate, a hysteresis loop is observed in the local spin valve, which is ascribed to the spin filtered transport.

\begin{figure}[t]
\centering
\includegraphics[scale=0.8]{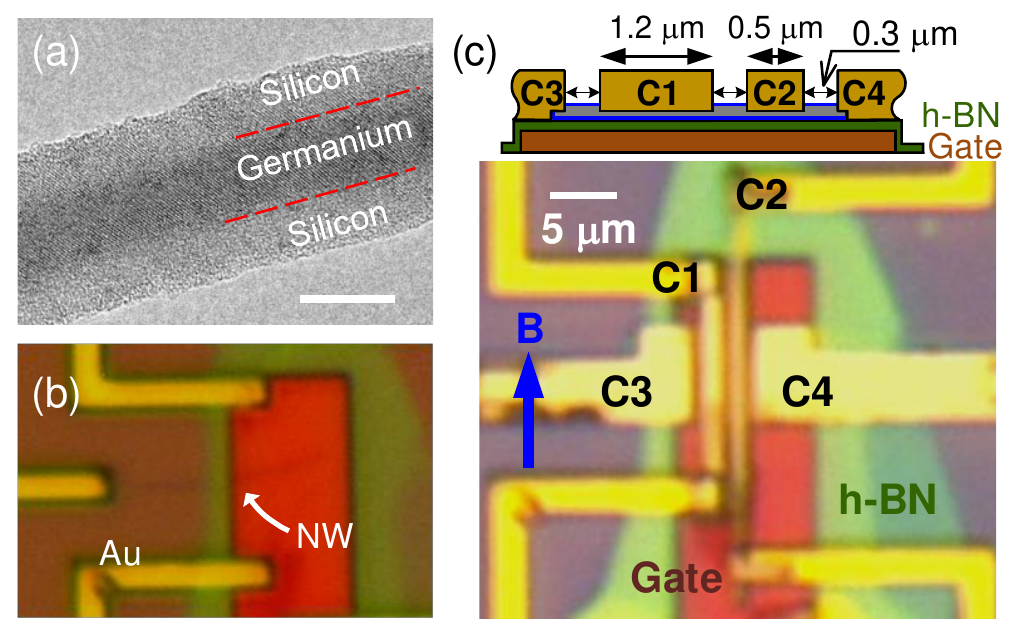}
\caption{(a) HRTEM image of a Ge/Si core/shell nanowire with $7\,$nm single crystalline Ge core in diameter and $5\,$nm-thick Si shell. Clear interface between Ge and Si is noted and highlighted by the red dashed lines. Scale bar: $10\,$nm. (b) \htc{Optical microscopy photo of a nanowire transferred on the h-BN flake covered pre-defined-gate, which is identified as a dark line.} Pre-defined gold contact electrodes are also shown. (c) \htc{Optical microscopy photo of a fabricated lateral spin valve device of Ge/Si core/shell nanowire.} Scale bar: $5\,\mu$m. Upper panel: Schematics showing the key dimensions of the device.} 
\label{fig:fig1}
\end{figure}

Epitaxial Ge/Si core/shell NWs were synthesized by a two-step vapor-liquid-solid method which has been described elsewhere.\cite{Yao-PNAS} The typical Ge/Si NWs used in this work have a single crystalline germanium core of $\sim 7\,$nm in diameter and $\sim 5\,$nm-thick silicon shell. Figure \ref{fig:fig1}(a) shows an example high resolution transmission electron microscope (HRTEM) image of one NW. The interface between Ge core and Si shell can be clearly identified from the electron transmission contrast. The relatively rough surface is due to the amorphous native silicon oxide formed naturally.
Device fabrication starts from dry-transferring a commercial available h-BN flake (Momentive, Polarthem grade PT110) of $\sim 30\,$nm-thickness onto a $15\,$nm pre-defined gold gate on a SiO$_2$/Si substrate using a home-made mechanical manipulator with a viscoelastic membrane (Gelfilm, Gelpak) \cite{Zomer-APL-2014, 2Dtrans}. 
Gold electrodes and pads for wire boding are defined from Ti/Au ($10\,$nm/$60\,$nm) using e-beam lithograph and evaporation.
Then, the NW was transferred onto the h-BN using a PMMA stamping technique described in detail elsewhere.\cite{Wang-APL-2016} Figure \ref{fig:fig1}(b) shows a transferred NW on h-BN/gate with gold electrodes nearby.
Finally, four $80\,$nm-thick cobalt ferromagnetic (FM) contacts with varied widths were deposited on the NW. A $20\,$nm-thick gold capping layer is used to protect cobalt from oxidation. Before evaporation, a short dip in buffered hydrofluoric acid is carried out to strip the thick natural oxide from the surface of the NW. 
The different widths of the FM contacts ensure that their
magnetization will be reversed at different $B$ fields. Larger width giving a lower exchange energy barrier and hence a smaller coercive field.

Figure \ref{fig:fig1}(c) shows one as-fabricated NW device. The key geometric parameters are indicated in the upper cross-section schematic.
Short NW channels of $300\,$nm are defined between four contacts.
The inner cobalt contacts C1 and C2 having the widths of $1.2\,\mu$m
and $0.5\,\mu$m are designed for the spin injection and detection, respectively. The outer two wide contacts having the width $> 5\,\mu$m are expected to have negligibly low coercive field compared to the inner two.
Measurements were performed in a pumped He-4 refrigerator at $1.5\,$K. The device is mounted on the sample insert with the long axes of its FM contacts aligned with the external $B$ field as illustrated in Fig. \ref{fig:fig1}(c). 
The differential conductance $G$ was measured using standard lock-in techniques with a low frequency of $74.7\,$Hz.

\begin{figure}[t]
\centering
\includegraphics[scale=0.65]{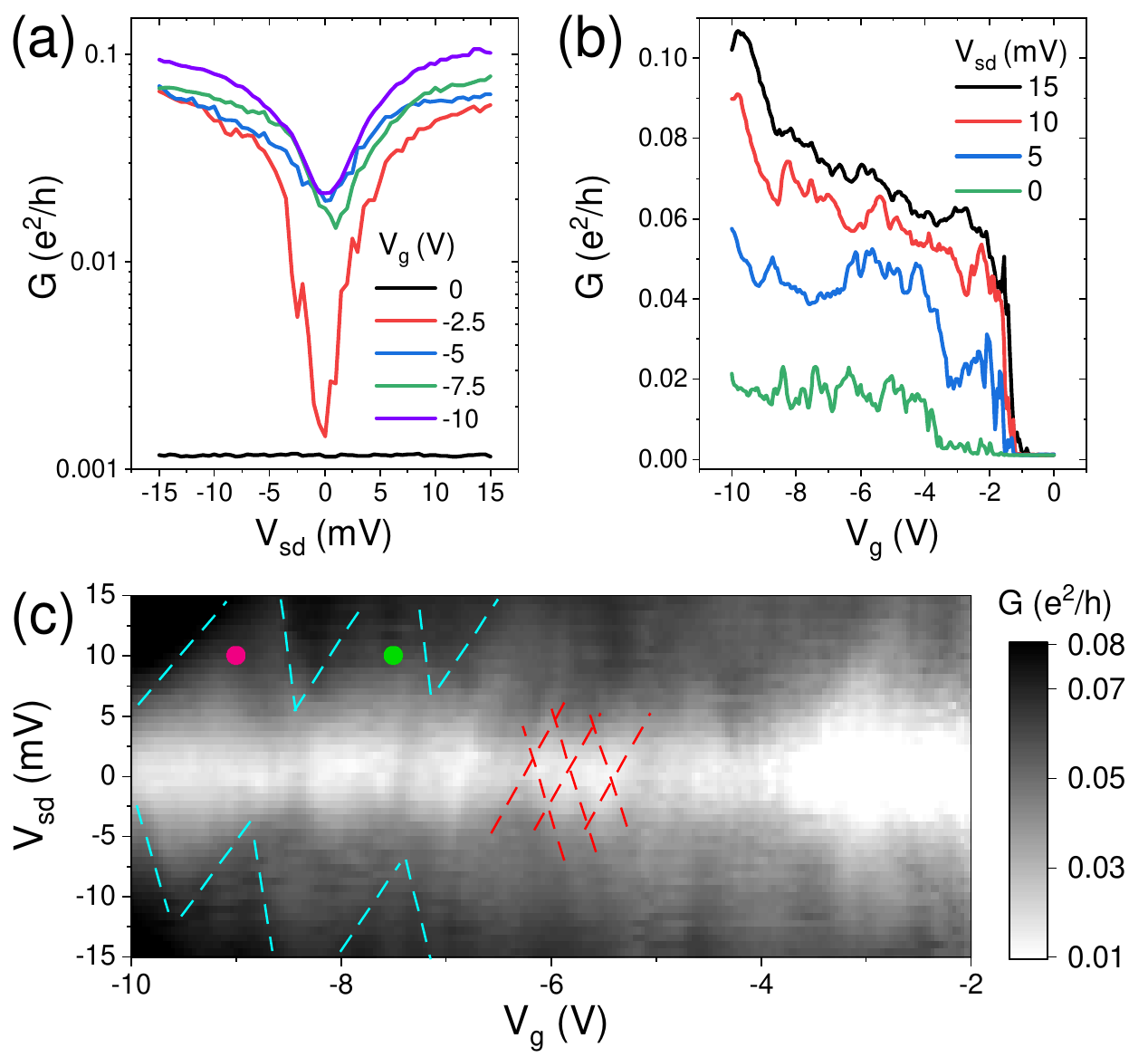}
\caption{(a) Differential conductance $G$ versus dc bias $V_{sd}$ between terminals 3 and 4 with various gate voltages $V_g$. (b) Gate dependence of differential conductance $G$ at varied dc biases $V_{sd}$. (c) Charge stability diagram showing the differential conductance $G$ measured under varied $V_g$ and $V_{sd}$. Lozenge patterns are noted as the signature of Fabry-P\'erot interference, which are highlighted by the red dashed lines. Light blue dashes indicate the identified perimeters of the blurry diamonds. Red and green dots indicate the measured points at bias of $10\,$mV with the gate of $-9\,$V and $-7.5\,$V, respectively.} 
\label{fig:fig2}
\end{figure}

It is known that a huge conductivity mismatch exists between metal contacts and semiconductors, which makes it challenging to detect the spin-polarization of electrons/holes flowing across a typical contact.\cite{Schmidt-PRB-2000} One of the commonly employed approaches to
address this issue requires a tunnel barrier made using molecular beam epitaxy or atomic layer deposited thin insulating layers at the contact interface, which is technically challenge especially on a nanowire.\cite{Tang-Nanoscale, Zhang-NL}
Here we demonstrate a thin natural oxide formed on the NW by a  controlled air exposure after HF etching could act as a tunnel barrier for spin injection.
Figure \ref{fig:fig2}(a) shows the differential conductance measured between C1 and C2 as a function of dc bias at various gate voltages. 
The non-linearity highlights the tunnelling nature of the contacts, while the symmetric curve reveals that the identical barriers are induced at the Co/NW interfaces by the natural oxidation.
Figure \ref{fig:fig2}(b) plots the gate dependence of differential conductance. The $p$-doping characteristic is noted by the enhanced conductance with negatively ramped gate voltage.
Compared to the ohmic contacted $300\,$nm-long NWs reported previously,\cite{SUN-NL} 
the conductance is one order of magnitude lower due to the existence of the tunnel barrier. \htc{We emphasize that a low spin injection efficiency is anticipated with this suboptimal tunnel barrier.}

\begin{figure}[t]
\centering
\includegraphics[scale=0.65]{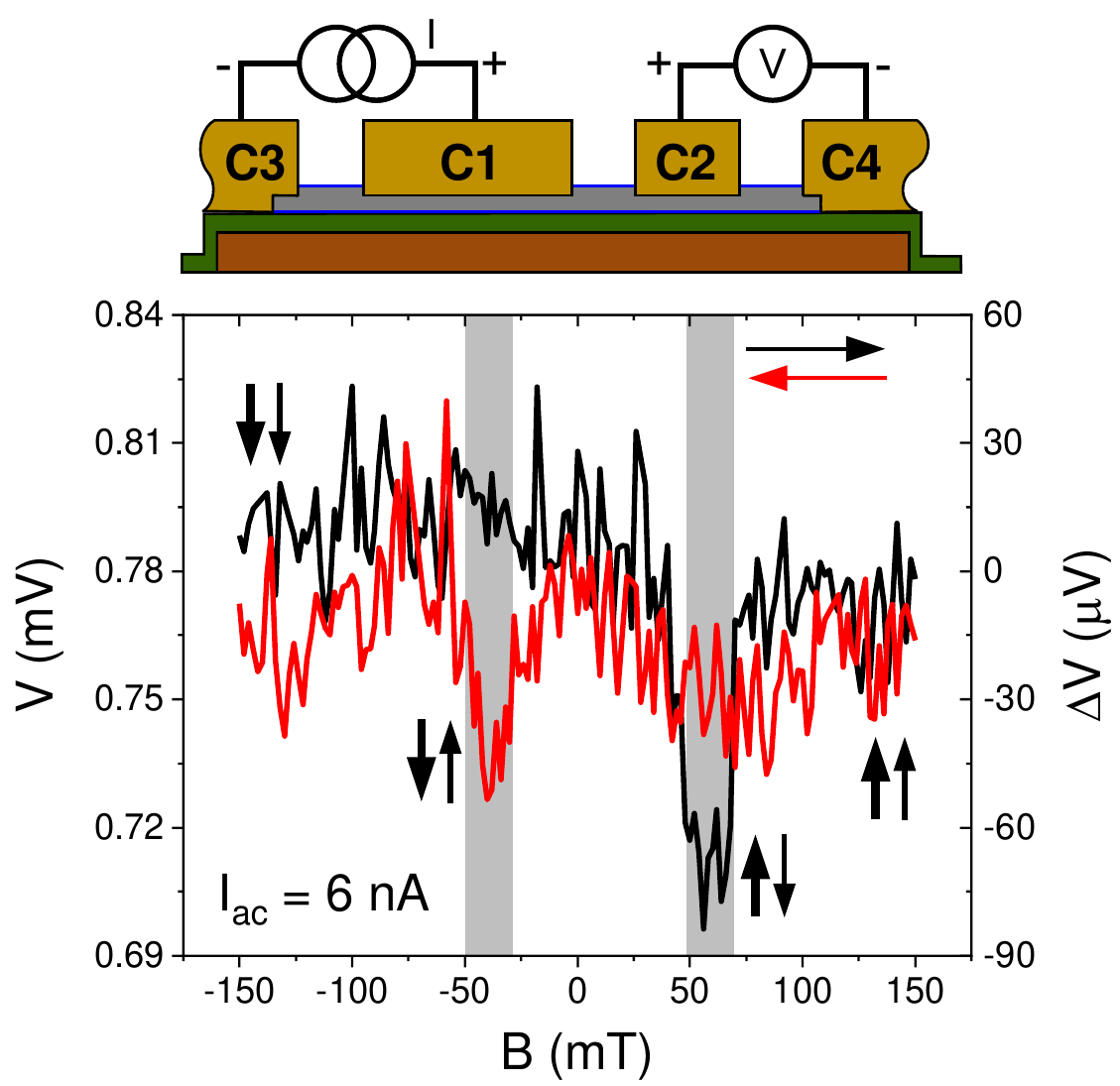}
\caption{Voltage output of the nonlocal measurements as a function of the in-plane $B$ field scanned in both up and down directions. The shadowed regions indicate the field range of the antiparallel magnetization in C1 and C2. Gate voltage is applied at $-10\,$V. \htc{Ac excitation current is $6\,$nA at $74.7\,$Hz.} The right y-axis indicates the relative voltage change. Upper schematic shows the nonlocal measurement configuration using a four terminal geometry.} 
\label{fig:fig3}
\end{figure}

Periodic oscillations are found superimposed on the transport curves, especially at zero bias. 
To understand the origin of the oscillations, we measure the charge stability diagram by scanning both gate voltage and dc bias (Fig. \ref{fig:fig2}(c)). The oscillations forming the lozenge shapes near zero bias resemble Fabry-P\'erot interference.\cite{Li-SR} 
Previously, the ``diamonds'' corresponding to quantized conductance plateaux and helical gaps have been observed in the charge stability diagram measured in the ohmic contacted $300\,$nm-long NWs.\cite{SUN-NL}
Also considering the long mean free path of $> 500\,$nm of the NW, we infer the ballistic transport is still realized in this $300\,$nm-long short NW junction.
\htc{However, the FM contacts with tunnel barriers required for spin injection hinders the clear measurements of quantized conductance and helical state, for which high quality ohmic contacts are necessary.}
With these Schottky contacts, the quantized conductance plateaux are inevitably smeared out heavily by the high resistance background and severe oscillations.\cite{Rainis-PRB}
Hence, when holes propagate phase coherently in the NW, they experience multiple partial reflections at the contact interfaces, therefore leading to the Fabry-P\'erot oscillations. 
We note that observation of a visually clear re-entrant conductance is not a pre-requisite to investigate the spin filtering effect. The spin-momentum locking is topologically protected anyway in the quasi-ballistic NW. 
Nevertheless, the re-entrant conductance feature if observable would make it easier to locate a helical gap. Otherwise, careful scanning of the gate voltage is compulsory to search for the distinct features originating from the helical gap.
\htc{An ohmic contacted NW device with similar geometry and h-BN dielectric in our previous work shows that a pseudo helical gap can be narrower than $1\,$V in the gate scan, making it difficult to locate.\cite{SUN-NL}}
Fortunately, three diamonds with the blurry perimeters can still be identified in Fig. \ref{fig:fig2}(c) at high negative gate voltages. \htc{(also see Fig. S1 in the supplementary material for the charge stability diagram with a different color scale.)}
As aforementioned, Ge/Si NWs possess a pseudo-helical gap in the same energy scale of quantized conductance at zero magnetic field.\cite{SUN-NL} Theoretically, every two adjacent conductance diamonds in the charge stability diagram should correspond to a pseudo-gap and trivial quantized conduction mode, respectively.\cite{Delft-helical, SUN-NL} 
\htc{Knowing the voltage separation of $5\,$V $\sim 8\,$V between the first conduction mode and pinch off from our previous work,\cite{SUN-NL} we speculate there is only one conduction mode and its associated helical gap within the applied gate range.}

We first confirm that polarized spins are injected at the cobalt FM contacts by measuring the spin valve effect in the NW. We carry out nonlocal measurements using a four-terminal configuration, which can effectively eliminate other effects showing similar features in the spin transport, \textit{e.g.} magneto-Coulomb effect and local Hall effects.\cite{nonlocal}
\htc{A $6\,$nA ac current excitation at the frequency of $74.7\,$Hz is applied
between C1 and C3.} A constant gate voltage is fixed at $-10\,$V. If spin-polarized holes are injected through C1 and accumulate in the NW. Subsequently, a net hole spin flow diffuses to the other side of the NW from C1 to C2, which can be detected as a voltage signal between the spin detector C2 and a remote contact C4.
The measured voltage is plotted as a function of $B$ in Fig. \ref{fig:fig3}. 
At $-150\,$mT, all FM contacts are magnetized toward
the negative direction. A nonzero background signal independent of the spin valve effect is 
observed, which is typically seen in the nonlocal measurements and is likely ascribed to the nonuniform electrical current injection at the contacts. 
As $B$ is ramped up to $50\,$mT, magnetization in the wide C1 contact flips. The antiparallel alignment causes a voltage drop of $\sim 60\,\mu$V, since the major spins in C2 have an opposite polarization to the ones accumulated in the NW. 
At a larger $B$ field of $70\,$mT, all FM contacts have their magnetizations 
parallelly aligned again. Then,the output returns to the background. A reverse scan shows similar behaviour with the anti-parallel magnetization between $-30\,$mT and $-50\,$mT. The slight asymmetry in two scans suggests a residual field of $10\,$mT in the external superconducting coil. \htc{Nonlocal spin valve effect is also measured at varied gate voltages and in a second device as shown in Figs. S2 and S3 in the supplementary material.}

\begin{figure}[t]
\centering
\includegraphics[scale=0.65]{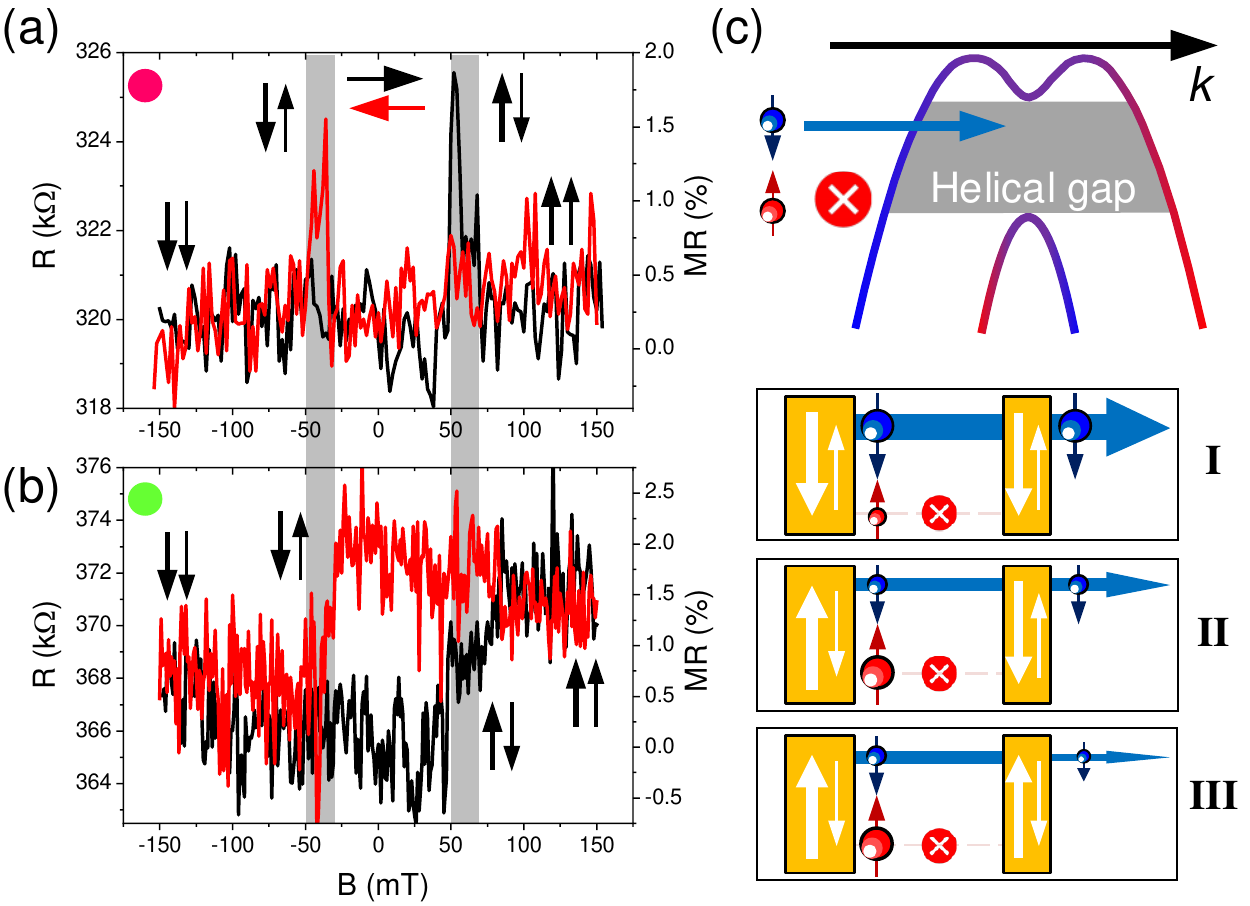}
\caption{Magnetoresistance measured between C1 and C2 with a constant dc bias of $10\,$mV at (a) $V_{\rm g} = -9\,$V (red dot in Fig. \ref{fig:fig2}(c)) and (b) $V_{\rm g} = -7.5\,$V (green dot in Fig. \ref{fig:fig2}(c)). The shadowed regions indicate the field range of the antiparallel magnetization in C1 and C2. The right y-axes show the MR ratio. (c) upper: schematic showing the spin filtering in the helical gap with a fixed momentum direction $k$. \htc{lower: illustrations of three resistance states \textbf{I}, \textbf{II}, and \textbf{III} in the up scan of $B$ when the channel is in helical gap (grey area in the above energy diagram) with a right-pointing momentum. The wide arrows in the contacts indicate polarization direction as well as the majority spin orientation, while the narrow arrows indicate the minority spin due to the low spin injection efficiency. The large and small size spin cartoons denote the major and minor spin injected at left FM contact, respectively. The thickness of the arrows indicates the strength of the corresponding spin flow.}} 
\label{fig:fig4}
\end{figure}

We now show the magnetoresistance measurements in two-terminal configuration between C1 and C2.
Two different gate voltages of $-9\,$V and $-7.5\,$V are applied to set the NW inside the left-most two adjacent diamonds highlighted in Fig. \ref{fig:fig2}(c).
A dc bias of $10\,$mV is applied in addition to an ac excitation of $20\,\mu$V, which helps to enhance the electrical signal and, more importantly, fixes the hole momentum pointing constantly from C1 to C2.
Figure \ref{fig:fig4}(a) and (b) present the results of these two measurements.
\htc{Large backgrounds are found in both measurements, which mainly originate from the Schottky contact with natural oxide as barrier.}
A difference of $0.37\times 2e^2/h$ between two background signals is consistent with the conductance re-entrant of the helical gap considering a finite temperature and superimposed Fabry-P\'erot oscillations.\cite{Rainis-PRB, SUN-NL}

More interestingly, two completely distinct magnetic field dependences are noted.
The one measured at $V_{\rm g} = -9\,$V exhibits the typical peak-like local spin valve signal. Due to the parallel and anti-parallel magnetizations in the two FM contacts, low and high resistance states are observed with their transition fields consistent with these found in the nonlocal measurements. 
This strongly suggests that these peak-like features observed in two-terminal configuration originates from spin transport in the Ge/Si NW. Furthermore, we understand that the gate voltage of $-9\,$V sets the NW in a trivial state outside the helical gap without selective spin transport.

On the contrary, a hysteresis loop with three resistance states is measured at $V_{\rm g} = -7.5\,$V. It can be explained by the spin filtering effect of the helical gap.
Fig. \ref{fig:fig4}(c) illustrates the scenario of the spin transport inside the helical gap with a low spin injection efficiency $\ll 100\%$.
Application of dc bias fixes the momentum direction in the NW, leading to a spin selective transport. Here we consider the situation that only the down-spins are allowed to transport in the NW. Although up-spins can still be injected, they are fully blocked in the channel by spin-momentum locking.
\htc{Scanning the $B$ field upward from its lowest value, a low resistance state is first observed when both FM contacts are magnetized parallel in the negative $B$ field. 
Down-spins are the major spins injected by C1 and can transport between two contacts with a low resistance as illustrated by state {\bf I} in Fig. \ref{fig:fig4}(c). 
As $B$ is ramped up to $\sim +50\,$mT, the magnetization flips in C1. 
Subsequently, the channel allowed down-spins become the minority. A relatively lower current than the previous parallel state is measured, giving the higher resistance (state {\bf II}).
When $B$ reaches $+70\,$mT, the magnetization in
C2 flips. This results in the parallel magnetization of the FM contacts again. However, the magnetization in C2 contact is opposite to the spin direction allowed in the channel (state {\bf III}). Subsequently, the resistance is further enlarged. }
In the reserve scan, these three resistance states are measured with their transitions at $-30\,$mT and $-50\,$mT, where the magnetizations in the two FM contacts are flipped, respectively.
\htc{Knowing that the pseudo helical gap is $\sim 0.5\,$V wide in the gate scan from our previous report,\cite{SUN-NL} the hysteresis loop in the magnetoresistance measurement indicating spin filtering is anticipated at different gate voltages close to $-7.5\,$V. We carry out another magnetoresistance measurement at $V_g$ of $-7.7\,$V, where the hysteresis loop is again observed (see Fig. S4 in the supplementary material). More importantly, background resistance and relative resistance changes at parallel and antiparallel polarizations in the contacts are consistent with those presented in Fig. \ref{fig:fig3}(b), verifying the spin filtering nature of the signal.} 
\htc{It is worth mentioning that the hysteresis loop signal of the spin filtering effect is not observed in the nonlocal measurement. Fig. S5 in the supplementary material shows the magnetoresistance measured with the nonlocal configuration at $V_{\rm g}$ of $-7.5\,$V. There, spins injected at C1 carry opposite momentum when flowing to the left and right sides of the NW. When the NW is set inside the helical gap, spin transport is always forbidden on one side of the NW at the contact due to spin-momentum locking. Consequently, no voltage drop is anticipated between C2 and C4 regardless the polarizations in the FM contacts.}

In summary, we have experimentally demonstrated the spin filtering effect originating from the helical state with spin-momentum locking in the quasi-ballistic Ge/Si core/shell nanowire devices. 
In the device, spin injection is simply realized by cobalt ferromagnetic contacts using the thin natural oxide on NW as tunnel barrier.
With a low spin injection efficiency, a hysteresis loop is measured as a function of $B$ field, when the channel is prepared in the helical gap and allows the transport of one polarized spin.
By electrically tuning the NW outside the helical gap, spin valve effect was observed instead in both local and nonlocal measurements. In the future, interface engineering can be used to significantly improve spin injection efficiency. On that basis, we propose a spin based transistor. Its ``OFF'' state is defined when injected polarized spin is fully blocked by the properly prepared the helical gap, while the spin transport can be turned ``ON'' by tuning the channel outside the helical gap electrically using a gate.\\
 
See the supplementary material for charge stability diagram, nonlocal spin valve measurements at varied gate voltages and in a different device, spin filtering effect measured with local configuration at gate voltage of $-7.7\,$V, and nonlocal measurement inside helical gap at gate voltage of $-7.5\,$V.\\

\htc{Note added: During the preparation of this manuscript, we noticed a recently published local spin valve study on InSb nanowires, which reports a spin filtering effect with similar appearance.\cite{InSb-SFE} }

\begin{acknowledgements}
The authors acknowledge Prof. C. M. Lieber for his support on the nanowire growth. This work was supported by National Natural Science Foundation of China (Grant No. 11804397), Hunan High-level Talent Program (Grant No.2019RS1006) and JSPS Grant-in-Aid for Scientific Research (B) (No. 19H02548).
\end{acknowledgements}

\section*{Data Availability}
The data that support the findings of this study are available from the corresponding author upon reasonable request.

\newpage



\begin{thebibliography}{[1]}
\bibitem{Xiang-Nature-2006}
H. Yan, J.  Xiang, W. Lu, Y. Hu, Y. Wu, and C. M. Lieber, C. M. \textit{Nature} \textbf{2006}, 441(7092), 489.

\bibitem{Lu-PNAS-2005}
W. Lu, J. Xiang, B.P. Timko, Y. Wu, and C.M. Lieber, Proc. Natl. Acad. Sci. U.S.A. \textbf{102}, 10046 (2005).

\bibitem{Li-APL}
J. Li, N. Jomaa, Y.-M. Niquet, M. Said, and C. Delerue, Appl. Phys. Lett. \textbf{105}, 233104 (2014).


\bibitem{Hu-NN-2012}
Y. Hu, F. Kuemmeth, C.M. Lieber, and C.M. Marcus, Nature Nanotechnology \textbf{7}, 47 (2012).

\bibitem{Fischer-PRL}
J. Fischer and D. Loss, Phys. Rev. Lett. \textbf{105}, 266603 (2010).
 
\bibitem{Brunner-science}
D. Brunner, B.D. Gerardot, P.A. Dalgarno, G. W\"ust, K. Karrai, N.G. Stoltz, P.M. Petroff, and R.J. Warburton, Science \textbf{325}, 70 (2009).

\bibitem{Loss-SOI}
C. Kloeffel, M. Trif, and D. Loss, Phys. Rev. B \textbf{2011}, 84(19), 195314.

\bibitem{Quay-NP-2010}
C. H. L. Quay, T. L. Hughes, J. A. Sulpizio, L. N. Pfeiffer, K. W. Baldwin, K. W. West, D. Goldhaber-Gordon, and R. de. Picciotto, Nat. Phys. \textbf{2010}, 6(5), 336.

\bibitem{Delft-helical}
J. Kammhuber, M.C. Cassidy, F. Pei, M.P. Nowak, A. Vuik, \"O. G\"ul, D. Car, S.R. Plissard, E.P. a. M. Bakkers, M. Wimmer, and L.P. Kouwenhoven, Nature Communications \textbf{8}, 1 (2017).

\bibitem{Julich-helical} 
S. Heedt, N. Traverso Ziani, F. Cr\'epin, W. Prost, S. Trellenkamp, J. Schubert, D. Gr\"utzmacher, B. Trauzettel, and T. Sch\"apers, Nature Physics \textbf{13}, 563 (2017).
 
\bibitem{SUN-NL}
J. Sun, R.S. Deacon, R. Wang, J. Yao, C.M. Lieber, and K. Ishibashi, Nano Lett. \textbf{18}, 6144 (2018).

\bibitem{TI-SF-1}
J. Liu, T.H. Hsieh, P. Wei, W. Duan, J. Moodera, and L. Fu, Nature Materials \textbf{13}, 178 (2014).

\bibitem{TI-SF-2}
S. Rachel and M. Ezawa, Phys. Rev. B \textbf{89}, 195303 (2014).

\bibitem{Yao-PNAS}
J. Yao, H. Yan, S. Das, J.F. Klemic, J.C. Ellenbogen, and C.M. Lieber, Proc. Natl. Acad. Sci. U.S.A. \textbf{111}, 2431 (2014).

\bibitem{Zomer-APL-2014}
P.J. Zomer, M.H.D. Guimar\~aes, J.C. Brant, N. Tombros, and B.J. van Wees, Appl. Phys. Lett. \textbf{105}, 013101 (2014).

\bibitem{2Dtrans}
A. Castellanos-G\'omez, M. Buscema, R. Molenaar, V. Singh, L. Janssen, H. S. J. Zant, and G. Ander Steele, 2D Mater. \textbf{1}, 011002 (2014)

\bibitem{Wang-APL-2016}
R. Wang, R.S. Deacon, D. Car, E.P. a. M. Bakkers, and K. Ishibashi, Appl. Phys. Lett. \textbf{108}, 203502 (2016).

\bibitem{Schmidt-PRB-2000}
G. Schmidt, D. Ferrand, L.W. Molenkamp, A.T. Filip, and B.J. van Wees, Phys. Rev. B \textbf{62}, R4790(R) (2000).

\bibitem{Tang-Nanoscale} 
J. Tang and K.L. Wang, Nanoscale \textbf{7}, 4325 (2015).

\bibitem{Zhang-NL} 
S. Zhang, S.A. Dayeh, Y. Li, S.A. Crooker, D.L. Smith, and S.T. Picraux, Nano Lett. \textbf{13}, 430 (2013).

\bibitem{Li-SR} 
S. Li, N. Kang, D.X. Fan, L.B. Wang, Y.Q. Huang, P. Caroff, and H.Q. Xu, Scientific Reports \textbf{6}, 1 (2016).

\bibitem{Rainis-PRB} 
D. Rainis and D. Loss, Phys. Rev. B \textbf{90}, 235415 (2014).

\bibitem{nonlocal} 
H. Idzuchi, Y. Fukuma, and Y. Otani, Physica E: Low-Dimensional Systems and Nanostructures \textbf{68}, 239 (2015).


\bibitem{InSb-SFE} 
Z. Yang, B. Heischmidt, S. Gazibegovic, G. Badawy, D. Car, P. A. Crowell, E. P.A.M. Bakkers, and V. S. Pribiag, Nano Lett. \textbf{20}, 3232 (2020).



\end{thebibliography}
\end{document}